\pdfoutput=1
\documentclass[aps,prd,amsmath,floats,floatfix,twocolumn,10pt,
superscriptaddress,nofootinbib,showpacs]{revtex4-1}

\usepackage[T1]{fontenc}
\usepackage[utf8]{inputenc}
\usepackage{lmodern}
\usepackage{verbatim}

\usepackage[dvipsnames, usenames]{xcolor}
\definecolor{linkcolor}{rgb}{0.0,0.3,0.5}
\usepackage[hypertexnames=false, unicode, colorlinks=true, linkcolor=linkcolor,
citecolor=linkcolor, filecolor=linkcolor,urlcolor=linkcolor,
pdfusetitle]{hyperref}

\usepackage[all]{hypcap}
\usepackage{graphicx}
\usepackage{xspace}
\usepackage{amssymb}

\usepackage{microtype}

\usepackage[english]{babel}
\usepackage{blindtext}

\usepackage[normalem]{ulem} 
\usepackage{bm} 
\usepackage{mathrsfs}

\usepackage{float}
\restylefloat{table}

\newcommand\eea{\end{eqnarray}}
\newcommand\bea{\begin{eqnarray}}
\newcommand\ea{\end{align}}
\newcommand\ba{\begin{align}}

\newcommand\nn{\nonumber}
\newcommand\ml{\mathscr}

\usepackage{natbib}
\usepackage{graphicx}
\usepackage[dvipsnames]{xcolor}
\usepackage{amsmath}
\usepackage{epsfig}

\begin{document}

\title{Numerical Evidence for Non-Axisymmetric Gravitational ``Hair'' for Extremal Kerr Black Hole Spacetimes with Hyperboloidal Foliations}
\newcommand{\URI}{\affiliation{Department of Physics, 
    University of Rhode Island, Kingston, RI 02881, USA}}    
\newcommand{\URICCR}{\affiliation{Center for Computational Research, 
    University of Rhode Island, Kingston, RI 02881, USA}}    
\newcommand{\UMassDMath}{\affiliation{Department of Mathematics, 
    University of Massachusetts Dartmouth, 285 Old Westport Rd., North Dartmouth, MA 02747, USA}}    
\newcommand{\UMassDPhy}{\affiliation{Department of Physics, 
    University of Massachusetts Dartmouth, 285 Old Westport Rd., North Dartmouth, MA 02747, USA}}    
\newcommand{\UMassDCSCDR}{\affiliation{Center for Scientific Computing \& Data-Science Research, 
    University of Massachusetts Dartmouth, 285 Old Westport Rd., North Dartmouth, MA 02747, USA}}    

\author{Som Dev Bishoyi}
\email{sbishoyi@umassd.edu}
\UMassDMath
\UMassDCSCDR

\author{Subir Sabharwal}
\email{subir@uri.edu}
\URICCR

\author{Gaurav Khanna}
\email{gkhanna@uri.edu}
\URI
\UMassDPhy
\UMassDCSCDR
\URICCR

\date{\today}

\begin{abstract}
 Various generalizations of the scalar, axisymmetric ``horizon hair'' for extremal black holes have recently appeared in the literature.  In this paper, we propose an expression for a non-axisymmetric gravitational ``charge'' and its potentially observable imprint at a finite distance from the horizon (Ori-coefficient) in extremal Kerr black hole backgrounds. Using a hyperboloidal foliation, we offer strong and robust numerical evidence for the potential existence of this horizon hair and its properties. Specifically, we consider the time evolution of horizon penetrating, quadrupolar and (subdominant) octupolar gravitational perturbations with compact support on extremal Kerr (EK) spacetime. We do this by numerically solving the Teukolsky equation and determining the conserved charge values on the horizon and at a finite distance from the black hole. 
\end{abstract}

\maketitle

\section{Introduction}


Black holes can be solely characterized by the trifecta of three parameters: mass, charge, and spin. These parameters completely specify the fundamental attributes of black holes, regardless of the intricate details of their formation. This simplicity is elegantly encapsulated by Bekenstein's renowned ``no-hair'' theorem~\cite{bek72a,bek72b,bek72c}, formalizing the idea that the external characteristics of a black hole can be entirely described by these three parameters.

There have been several important black hole solutions with hair that have been found as exceptions to Bekenstein's theorem. Examples include colored black holes which are static solutions to the Einstein-Yang-Mills (EYM)~\cite{hbh-1,hbh-2,hbh-3,hbh-33} or Einstein-Yang-Mills-Higgs systems~\cite{hbh-6} yielding non-abelian hair. Additionally, there are black hole solutions with topological features like skyrmionic hair~\cite{hbh-4,hbh-5}. These hair typically are manifested as a result of non-linear field configurations outside the black hole that extend at least beyond the $3/2$ of the horizon radius for Schwarzschild black holes~\cite{hbh-9, hbh-10}. Some of these works have also been extended to the case of sub-extremal rotating black holes~\cite{hbh-7, hbh-8}.

Extremal black holes saturate geometric inequalities for the mass, angular momentum and charge~\cite{akk,dain}. These black holes play an important role in the study of Hawking radiation and more generally in quantum gravity~\cite{hhr}, and in string theory~\cite{sv} due to the fact that they have zero temperature. Their near-horizon geometry produces new solutions to the Einstein equations with conformally invariant properties~\cite{cdkktp}. Both gravitational and electromagnetic signatures of the near-horizon geometry have been introduced in~\cite{ghw,gls}. Astrophysical evidence supports the existence of both stellar near-extremal black holes~\cite{vmqr} and super-massive near-extremal black holes~\cite{bren}.

Recent investigations in the case of extremal black holes have revealed nuances to the seemingly comprehensive ``no-hair'' theorem. Pioneering work by Aretakis and collaborators~\cite{are12,aag18,lmrt13,zimm17,lr12,gz18,bk18,bks21,bks23,aks23} on extremal black holes, specifically focusing on extremal Reissner-Nordstr{\"o}m (ERN) and extremal Kerr (EK) black holes, suggests a potential ``loop hole'' in the context of linearized theory. Through rigorous analyses and numerical simulations, these works demonstrate that certain conserved charges persist on the event horizon (EH) of these black holes, challenging the traditional notion of black hole simplicity.

The conserved charge identified by Aretakis and his colleagues is evaluated solely based on the radial derivative of an axisymmetric scalar or gravitational field on the EH $\ml{H}^+$ at late times. This charge bears the hallmarks of a ``hair'' since it can be observed from the exterior, marking a departure from the expectations set by the ``no-hair'' theorem. Further analytical study in Ref.~\cite{aag18}, revealed that for extremal Reissner-Nordstr{\"o}m black holes, the horizon hair could be precisely calculated by monitoring the field values at null infinity, $\ml{I}^+$. 

Ori's contributions~\cite{ori} supplemented the understanding of Aretakis charges, providing an axisymmetric expression for these charges at finite radial distances from an extremal Reissner-Nordstr{\"o}m black hole. Building upon this groundwork, numerical simulations expanded the scope to include the case of extremal Kerr black holes, encompassing both scalar and gravitational field scenarios~\cite{bks21,bks23,aks23}. The numerical verification of the existence of horizon hair for extremal Reissner-Nordstr{\"o}m and extremal Kerr black holes, calculable at $\ml{H}^+$ and $\ml{I}^+$, underlines the potential of such results for additional computational study.

In this paper, we offer numerical evidence for conjecture that the notion of horizon hair extends to the non-axisymmetric gravitational perturbations as well. Precisely, these are modes i.e. coefficients in the expansion of the perturbation in the basis of spin-weighted spherical $_{s}Y_{lm}(\theta,\phi,\omega)$ harmonics having $m\neq0$. This is significant since non-axisymmetric, radiative,  gravitational field perturbations are precisely what gravitational wave observatories such as LIGO, Virgo and KAGRA are designed to detect. Our study marks the first attempt in bringing the subject of horizon hair -- primarily studied in mathematical relativity -- directly into the realm of observability. We will focus on the quadrupolar ($\ell = m = 2$) cases in this paper, since it  is the most dominant radiative multipole of the gravitational field. We also include a study of the sub-dominant octupolar ($\ell, m = 3, 2$) case. 

The proposed non-axisymmetric hair is based on the so-called Beetle-Burko scalar i.e. $\psi_0\psi_4$~\cite{beetle-burko-2002} where $\psi_0$ and $\psi_4$ are the Weyl scalars that capture the radiative degrees of freedom of the gravitational field. Details on this scalar appear later in this paper; for now, we simply comment that this Beetle-Burko scalar is an invariant with respect to both coordinate and the null tetrad transformations, and thus can be considered as a purely geometrical quantity encoding information about the physical curvature of the spacetime. In fact, the scalar is closely related to the Kretschmann scalar, as mentioned later this paper. {\em The proposed non-axisymmetric charge is defined as the magnitude of the transverse derivative of the Beetle-Burko scalar on $\ml{H}^+$ at late times.} We will demonstrate numerically that such a quantity is conserved for extremal Kerr black holes, at late times, while the scalar itself decays with an inverse power of time on $\ml{H}^+$. Note, that our proposed expression is only valid at late-times, so it should interpreted as the first term in the inverse-time expansion of a conjectured conserved charge. Finally, we will also show that this proposed charge is calculable from the extremal Kerr black hole's exterior, thus  offering numerical evidence for the potential existence of a non-axisymmetric, gravitational horizon hair. This exploration of extremal black holes, their conserved charges, and the intricate interplay between the interior and exterior dynamics challenges our conventional understanding of black hole simplicity in the context of the no-hair theorem.

To be very explicit, in this work: (1) we conjecture the existence of a non-axisymmetric gravitational charge and horizon hair in EK; (2) we perform a late-time study i.e. our results are relevant to the first-term in an inverse-time expansion of the quantities we consider; and (3) we provide numerical evidence for the first-term of an expression of the charge and horizon hair i.e. an approximation to the conjectured quantity.

Note that our work is closely related to the notion of ``Gajic instabilities''~\cite{gajic} that refer to unbounded growth of the transverse derivatives of the non-axisymmetric scalar field on the horizon of extremal Kerr black holes. However, the focus of this work is on the gravitational case. 

We do not conduct a study of near-extremal holes in this paper. However, our previous work~\cite{nek,nek2} in that context suggests that for cases in which the black holes are very close to extremality, the extremal case results are a good approximation as a ``transient'' i.e. for early times. Therefore, the results presented in this paper may be seen as applicable to the near-extremal cases as well, albeit for a short duration in time. In addition, we use an approximation that ignores important nonlinear effects in Einstein's theory which would likely result in the hair not lasting forever even in the extremal case. However, a plausible scenario for measurement may involve a rotating black hole spinning sufficiently fast allowing for the effects of the `transient hair' to have a measurable impact on a gravitational wave detector.

This paper is organized as follows: In Section II, we briefly summarize the details of the formulation, coordinate system and numerical approach -- adapted from our previously published works~\cite{bks21,bks23}. In Section III, we offer the conjectured expression for a new non-axisymmetric conserved charge and provide strong and compelling numerical evidence for it. That is followed by Section IV that demonstrates that the proposed charge can be computed externally to the extremal black hole. We end the paper with Section V that summarizes our key results.

\section{Ori's Late-time expansion, Coordinates and Numerical Details}
We borrow our notation from previous work~\cite{bks21,bks23} where the late time expansion of a field $\Psi$ in a black hole space-time was written as:
\bea\label{ori_expansion}
    \Psi_{s,\ell,m}(t,r) = && e_{s,\ell,m} r^{-q_{s,\ell,m}}(r-M)^{-p_{s,\ell,m} } t^{-n_{s,\ell,m} } \nn\\ 
     +&&\mathcal{O}(t^{-n_{s,\ell,m}-k_{s,\ell,m} })
\eea
in Boyer-Lindquist coordinates, where $s$ is the field's spin-weight, ($\ell, m$) characterize the spin-weighted spherical harmonic modes. $e_{s,\ell,m}$ is the so-called Ori-coefficient while $q_{s,\ell,m}$, $p_{s,\ell,m}$ and $n_{s,\ell,m}$ are power-law exponents for the various $(s,\ell,m)$. The second term on the right-hand-side refers to terms that decay faster in time than $t^{-n_{s,\ell,m}}$. This late-time expansion is expected to be valid for $t\gg |r_*|$, where $r_*$ is the tortoise coordinate. It is important to note that the decay rate $n_{s,\ell,m}$ depends critically on whether or not the initial data has support on the horizon. We compare the Ori-coefficient $e_{s,\ell,m}$ with the charge derived from the field on $\ml{H}^+$ and as done previously~\cite{bks21,bks23}, we will identify it as a ``hair'' since it is constant and in-principle observable externally.

To do this, we solve the Teukolsky equation~\cite{teuk1973} for perturbations in EK backgrounds using the Hawking-Hartle tetrad, focusing on non-axisymmetric modes ($\ell=2, 3; m=2$). We modify the equation to work in compactified hyperboloidal coordinates $(\tau, \rho, \theta, \phi)$ that allow for time evolution on hypersurfaces which bring $\ml{I}^+$ to a finite radial coordinate $\rho(\ml{I}^+)=S<\infty$. Instead of approaching spatial infinity as Cauchy surfaces do, they reach null infinity $\ml{I}^+$ , which makes them suitable for radiation extraction. The compactification allows us to extract signals at (future null) infinity ($\ml{I}^+$) directly, because it is part of the computational domain~\cite{z08}.

The relationship between these new coordinates $(\tau,\rho)$ and the spherical Boyer-Lindquist coordinates $(t,r)$ is~\cite{bkz16}
\bea\label{hypcoord}
\Omega(\rho) &=& 1-\frac{\rho}{S}\nn\\
r &=& \frac{\rho}{\Omega(\rho)}\\
v\equiv t+r_*-r &=& \tau+\frac{\rho}{\Omega(\rho)}-\rho-4M\log\Omega(\rho)\nn
\eea
where $S$ denotes the location of $\ml{I}^+$ in hyperboloidal coordinates, $r_*$ is the usual tortoise coordinate and $v$ is the modified advanced time. The free parameter S controls the numerical domain and the foliation, and gives us some freedom
in the number of grid points and the size of the time
steps~\cite{hbnz14}. In Eq.\eqref{hypcoord}, we map the unbounded domain $[1, \infty]$ in $r$ to the computational domain $[0.95,S]$ in $\rho$ (see below). The choice of $\Omega$ determines the properties of the compactification. Although there is freedom in the choice of $\Omega$, there are general properties that it must satisfy which can be found in~\cite{zk11}. Our hypersurfaces are different from the Boyer–Lindquist hypersurfaces in such a way that approaching $\ml{I}^+$ the spatial wavelength of outgoing radiation is unbounded in our foliation (“$\tau = \text{const}$ hypersurfaces are asymptotically wave fronts”). Therefore we can resolve the field on the compactified grid and do not encounter spurious reflections due to compactification near the outer grid boundary~\cite{z11}.

Our numerical implementation scheme entails re-writing the second order PDE in terms of two coupled first-order differential equations. We solve this system using a high-order WENO finite-difference scheme with explicit Shu-Osher time-stepping. Details may be found in our previous work~\cite{camc21}. We set the mass scale $M=1$. $S=19.0$ and the location of $\ml{H}^+$ such that for a similar mass extremal black hole, $\rho(\ml{H}^+)=0.95$. The initial data is a truncated Gaussian centered at $\rho=\{1.0, 1.1, 1.2, 1.3, 1.4\}$ with a width of $0.22$ and non-zero for $\rho\in[0.95,8]$. This ensures compactly supported initial data but with non-zero support on the $\ml{H}^+$ surface. 

{\em\underline{The importance of hyperboloidal compactification}:} Hyperboloidal slicing allows for a seamless connection between the black hole horizon and future null infinity. It provides a geometric regularization of time-harmonic oscillations, which are crucial for analyzing quasi-normal modes (QNMs)~\cite{rpm,jlr}. The hyperboloidal approach facilitates the use of pseudospectrum analysis, which is essential for understanding the spectral stability of black hole QNMs. This method has been applied to various spacetimes, including Schwarzschild and Reissner-Nordström black holes, revealing insights into the stability and instability of QNMs under perturbations~\cite{jms,dmbcj}.

For the purpose of this study, the use of hyperboloidal compactification played a critical role in the context of our computational capabilities. Not only does the method solve the so-called ``outer boundary problem'' allowing for arbitrarily long evolutions with no change to the computational domain, it also allows for much higher grid densities in the strong field because the compactification compresses the domain by several orders of magnitude ~\cite{bkz16}. That combination results in {\em higher accuracy} at {\em negative cost}! We could not perform the computations necessary for this study in the absence of hyperboloidal compactification. 

Despite that fact, it is worth noting that these simulations still take significant computational resources to run. Each simulation takes a month of wall-time using 8 Nvidia V100 GPUs each. For this reason, we focus our efforts on the extremal cases and do not attempt any near-extremal cases. However, our previous work~\cite{nek,nek2} in that context suggests that for cases in which the black holes are very close to extremality, the extremal case results are a good approximation as a ``transient'' i.e. for early times.

Note that we solve the Teukolsky equation, for both $s=\pm 2$. Recall that the solution for $s=+2$ corresponds to the Weyl scalar $\psi_0$ while that for $s=-2$ corresponds to the Weyl radiation scalar, $\psi_4$~\cite{teuk1973}.

\section{Non-Axisymmetric Gravitational Perturbations on Extremal Kerr}
The gravitational scenario under consideration exhibits Ricci flatness, implying that all scalars formulated with $R$ or $R_{\mu\nu}$, and their derivatives, vanish identically. Consequently, the curvature of spacetime relies solely on the Weyl tensor. In the context of 4-dimensional spacetime, Petrov~\cite{petrov} established that there exist fourteen algebraically-independent scalars crucial for determining curvature properties. However, when confined to vacuum, only four non-vanishing scalars persist~\cite{beetle-burko-2002}.

In our specific scenario, additional constraints arise, notably the fact that the Hawking-Hartle tetrad constitutes a transverse frame ($\psi_1=0=\psi_3$). Given the known EK spacetime background, specifically the constancy of the Weyl scalar $\psi_2$ along the event horizon $\ml{H}^+$ 
 ~\cite{Poisson:2004cw}, only two algebraically-independent curvature scalars endure. These can be identified as the real and imaginary segments of the Beetle-Burko scalar $\xi = \psi_0\psi_4$~\cite{beetle-burko-2002}, which exclusively encapsulates information pertaining to the radiation field in regions where gravitational radiation can be unambiguously defined~\cite{beetle-burko-2002}.

Our computational approach enables the determination of the Weyl scalars $\psi_0$ and $\psi_4$ within the Hartle-Hawking tetrad. It is noteworthy that while $\psi_4$ and $\psi_0$ behave as scalars under coordinate transformations, they lack invariance under tetrad transformations. In contrast, the product $\psi_0\psi_4$ not only retains scalar behavior under coordinate transformations but also maintains invariance under tetrad transformations~\cite{beetle-burko-2002,BBBN-2005}. It is this product of the two Weyl scalars that encapsulates information crucial for determining the physical properties of the spacetime under consideration. It is also worth pointing out that the Kretschmann scalar can be written in terms of the Weyl scalars as,
\bea
K=8(\psi_0\psi_4+3\psi_2{\,^2}-4\psi_1\psi_3)+{\rm c.c}
\eea
where ${\rm c.c.}$ is the complex conjugate of the first term~\cite{cbcr}. As the Hartle-Hawking frame is transverse ($\psi_1=0=\psi_3$), and the Weyl scalar $\psi_2$ maintains its Kerr background value within linearized theory, the Kretschmann scalar is very closely related to the Beetle-Burko scalar -- the latter can be thought of as a perturbation on the former's background value.

\begin{figure}[!ht]
    \centering
    \includegraphics[width=\columnwidth]{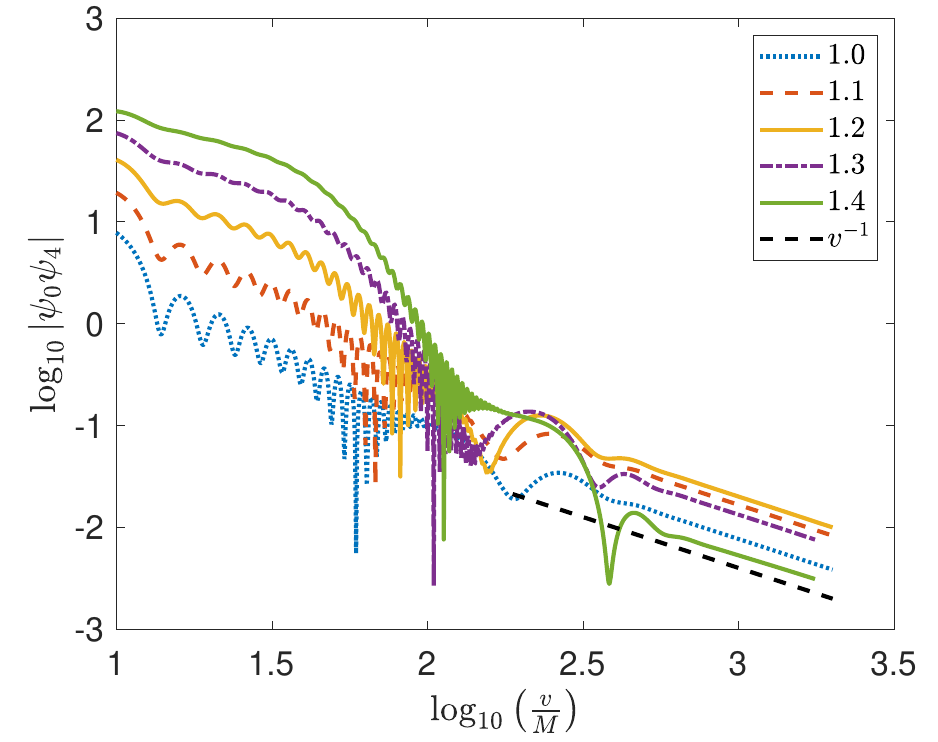}
    \caption{Plots of the Beetle-Burko scalar $|\xi_{2,2}|$ as function of advanced time $v$ on $\ml{H}^+$. $|\xi_{2,2}|$ decays with an inverse power of $v$. We show this for various initial conditions that are labeled by the centers of the respective Gaussians.}
    \label{fig:BBv}
\end{figure}

\begin{figure}[!ht]
    \centering
    \includegraphics[width=\columnwidth]{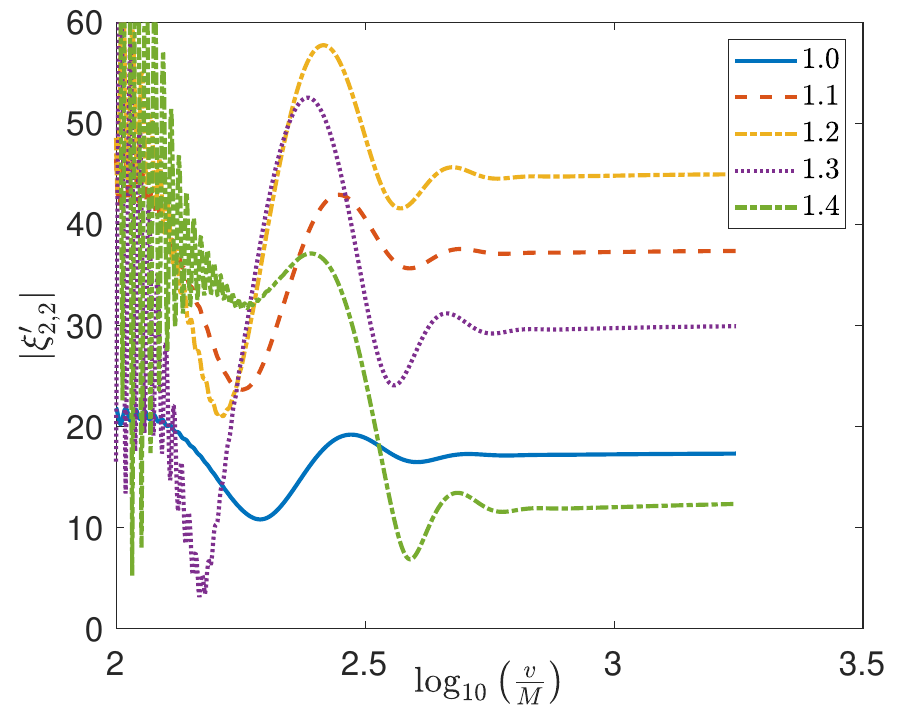}
    \caption{Plots of the proposed non-axisymmetric charge $|\xi'_{2,2}|$ on $\ml{H}^+$ as function of advanced time $v$. While $|\xi_{2,2}|$ decays with an inverse power of $v$, the proposed quadrupolar charge $|\xi'_{2,2}|$ is a constant at late times.}

    \label{fig:Hv}
\end{figure}

Seeking inspiration from definition and properties of the Beetle-Burko scalar, let us define a quantity $\xi_{\ell,m}$ for a particular multipole $\ell,m$ to be: 
\bea\label{xi}
    \xi_{\ell,m}(t,r) = \Psi_{2,\ell,m}(t,r) \; \Psi_{-2,\ell,m}(t,r) 
\eea
where $\Psi_{\pm2,\ell,m}(t,r,\theta)$ are the computed solutions to the Teukolsky equation for $s=\pm2$ and projected to the $\ell, m$ mode. More specifically, in this paper we choose the quadrupolar ($\ell=m=2$) and octupolar ($\ell, m = 3, 2$) modes since they are the well-known dominant radiative terms, and thus serve as reasonable approximations to $\xi$. In Figs.~\ref{fig:BBv},~\ref{fig:Hv} we show that while $|\xi_{2,2}|$ decays with inverse time $v$ on $\ml{H}^+$, our proposed quadrupolar charge $|\xi'_{2,2}|$ is conserved at late times. Here the prime refers to a transverse derivative i.e. with respect to $r$. {\em This offers evidence for a potential non-axisymmetric conserved charge for extremal Kerr black holes built using a geometric quantity that is coordinate and tetrad invariant}. Note, however, this quantity is conserved only at late-times not in the early part of the evolution. This implies that the proposed quantity is only the first term for a potential conserved charge in an inverse-time expansion, and thus should be interpreted as an approximation. 

\begin{figure}[!ht]
    \centering
    \includegraphics[width=\columnwidth]{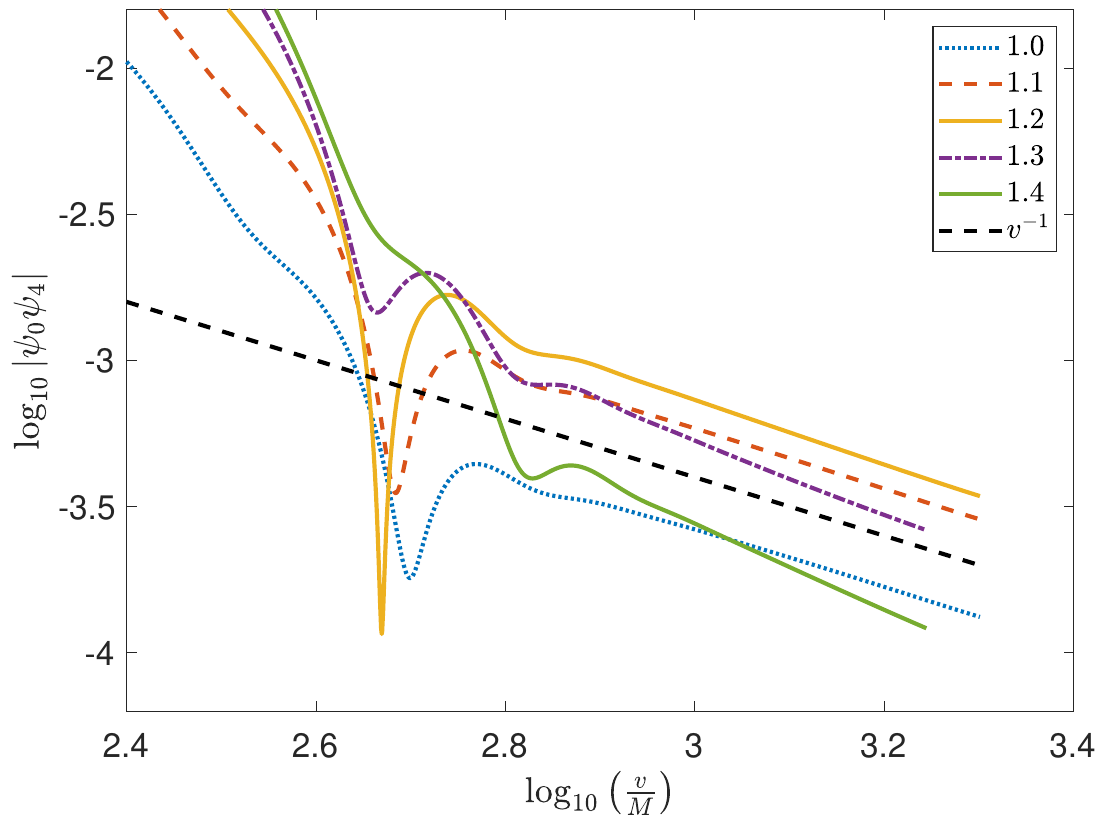}
    \caption{Plots of the Beetle-Burko scalar $|\xi_{3,2}|$ as function of advanced time $v$ on $\ml{H}^+$. $|\xi_{3,2}|$ decays with an inverse power of $v$.}
    \label{fig:BBv3}
\end{figure}

\begin{figure}[!ht]
    \centering
    \includegraphics[width=\columnwidth]{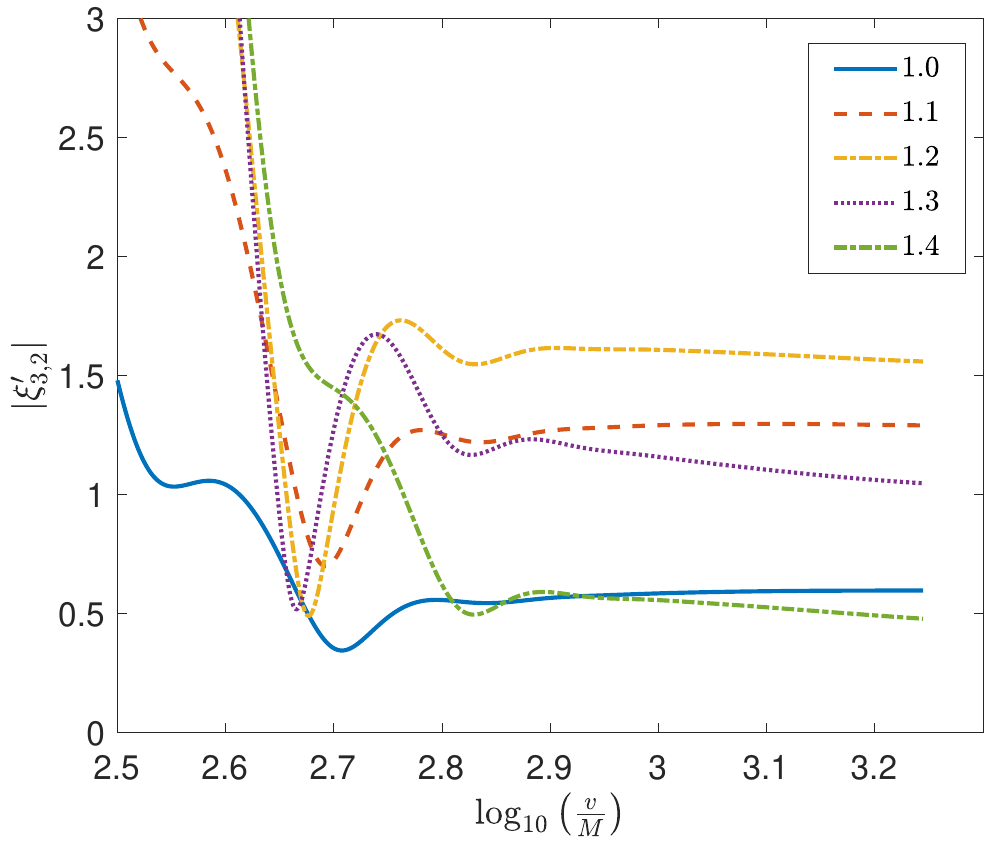}
    \caption{Plots of the proposed non-axisymmetric charge $|\xi'_{3,2}|$ on $\ml{H}^+$ as function of advanced time $v$. While $|\xi_{3,2}|$ decays with an inverse power of $v$, the proposed octupolar charge $|\xi'_{3,2}|$ is a constant at late times.}
    \label{fig:Hv3}
\end{figure}

In Figs.~\ref{fig:BBv3},~\ref{fig:Hv3} we show the same data for the octupolar ($\ell, m = 3, 2$) case. While those figures offer compelling evidence for a conserved octupolar charge, it is worth noting that for the cases with Gaussian center further away i.e. $1.3$ and $1.4$, we see that the data doesn't appear to have settled down to the asymptotic behavior for the duration of our computations. Nonetheless, we consider the data presented as evidence for the potential existence of an octupolar charge.     

Note that this result isn't entirely unexpected -- our results for $\psi_4$ and $\psi_0$ are in agreement with the results of Ref.~\cite{casals-2016,bk18} for $n=0,1,2,3$, i.e. the late time behavior is found to be $\psi_4^{(n)}(v\gg M)\sim v^{3/2+n}$ and $\psi_0^{(n)}(v\gg M)\sim v^{-5/2+n}$ where $n$ is the transverse derivative order and $v$ is the advanced time coordinate.  Combining those for $\xi_{2,2}$ yields a decay of $1/v$, while $|\xi'_{2,2}|$ yields a constant, exactly as we have uncovered numerically. 

We next proceed to demonstrate that this quantity can computed from outside the black hole in the bulk inspired by Ori's approach at a finite radial distance from the black hole's horizon~\cite{ori}. 

\section{Ori expansion for gravitational perturbations}
\begin{figure}[!ht]
    \centering
    \includegraphics[width=\columnwidth]{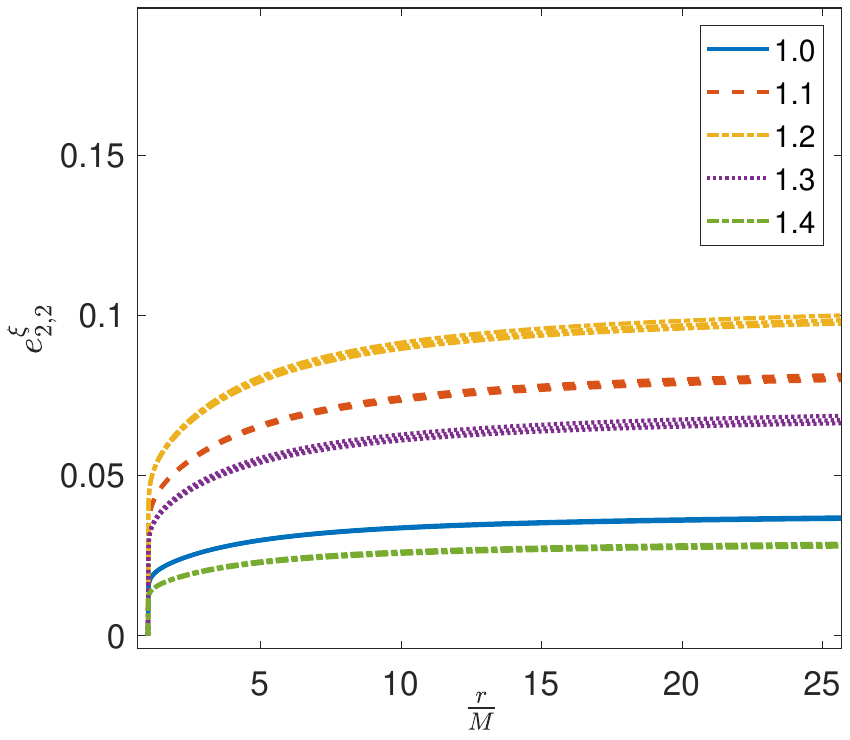}
    \caption{The values of $e_{2,2}^{\xi} = |\xi_{2,2}(t,r)|\;g(t, r)$ as functions of $r/M$ for extremal Kerr. These plots show values of the data set for which the Gaussian’s center is at $\rho/M = \{1.0, 1.1 ,1.2,1.3, 1.4\}$. The values are plotted for $t/M =$ 1000, 1200, 1400. The function $g(t, r) = (t/M)^{2}(r/M)^{-1}(r-M)$.}
    \label{fig:OriPhi_r}
\end{figure}

To extract the proposed charge data outside the black hole, we solve the Teukolsky equation~\cite{teuk1973} with $s=\pm 2$ (radiative fields) for $a=M$, where $a$ is the Kerr spin parameter of the black hole. Below, we concentrate on the non-axisymmetric mode $m=2$, and so the dynamical system to solve is $(2+1)$-D in $(\tau,\rho,\theta)$. The solution is projected down to $\ell=m=2$ and compared to the form: 
\bea\label{ori_expansion2}
    |\xi_{2,2}(t,r)| =&& e_{2,2}^{\xi} r^{-q_{2,2}^\xi}(r-M)^{-p_{2,2}^{\xi}} t^{-n_{2,2}^{\xi}}\nn\\
    +&&\mathcal{O}(t^{-n_{2,2}^{\xi}-k_{2,2}^{\xi}})
\eea
The form of Eq.(\ref{ori_expansion2}) has been constructed by combining Eq.~\eqref{ori_expansion} and Eq.~\eqref{xi} and rearranging powers of $r,r-M$ and $t$. To calculate the exponents $p$ and $q$ that appear in Eq.~\eqref{ori_expansion2}, we note that Ori expansion is expected to correctly model the $r$ dependence of the Beetle-Burko scalar in the large $t$ domain i.e $t\gg|r_{*}|$, where $r_{*}$ is the tortoise coordinate. Thus, we select the data in the intermediate region $r \approx 2 $ to $r \approx 20$ for fitting. Since large times are preferable, we use $t=1400$ when removing the time dependence from Eq.~\eqref{ori_expansion2} by multiplying $t^{n^{\xi}_{2,2}}$ on both sides. Since Fig.~\ref{fig:OriPhi_r} shows robustness for different time slices, any value of $t \sim \mathcal{O}(1500)$ can be chosen. The quantity $|\xi_{2,2}(t=1400,r)|\cdot t^{n^{\xi}_{2,2}}$ is fitted with the model $e_{2,2}^{\xi}r^{-q}(r-M)^{-p}$ where $M=1$.

\begin{figure}[!ht]
    \centering
    \includegraphics[width=\columnwidth]{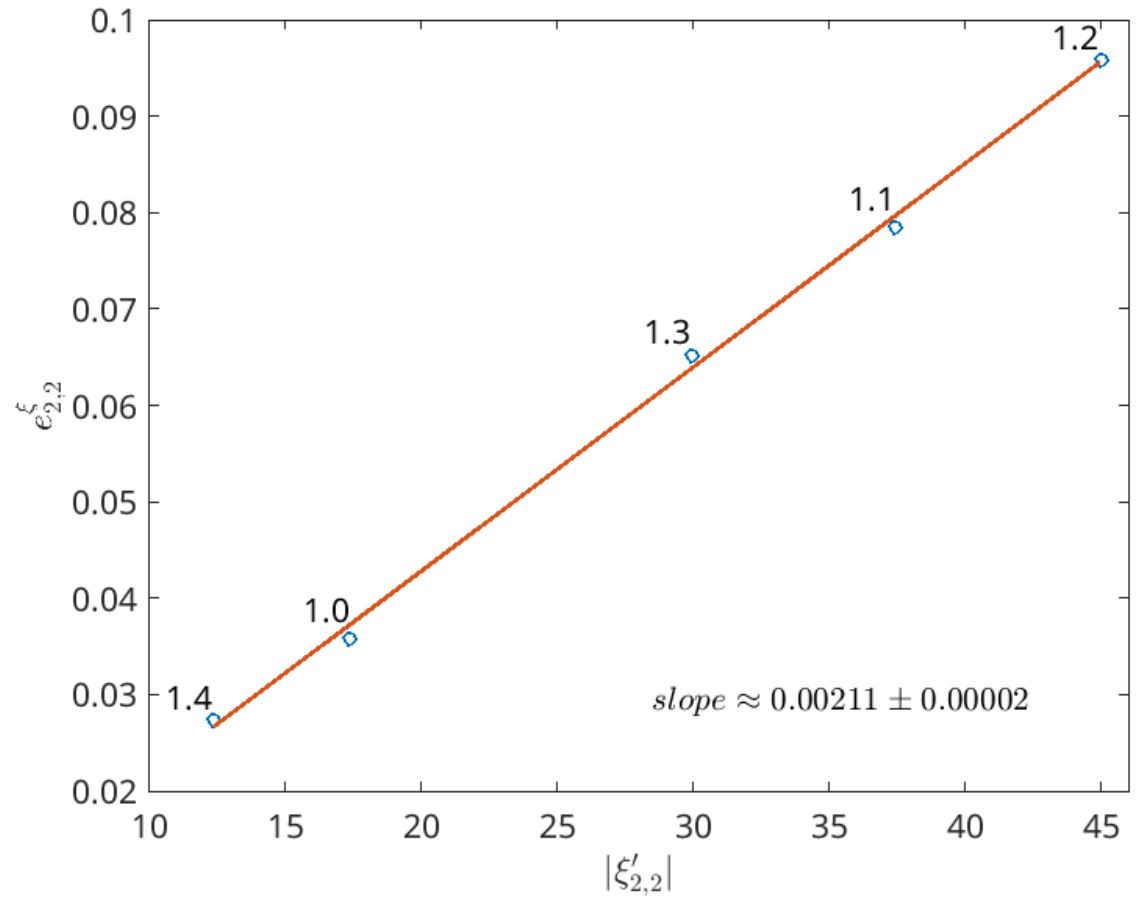}
    \caption{The values of $e_{2,2}^{\xi}$ plotted against the proposed charge $|\xi'_{2,2}|$. $e_{2,2}^{\xi}$ is computed using data external to the extremal Kerr black hole while $|\xi'_{2,2}|$ is measured on the $\ml{H}^+$. The plot depicts values of those quantities for which the Gaussian’s center is at $\rho/M = \{1.0, 1.1, 1.2, 1.3, 1.4\}$. The linear relationship between the two quantities explicitly demonstrates that one can compute $|\xi'_{2,2}|$ from $e_{2,2}^{\xi}$.
    }
    \label{fig:HvsE}
\end{figure}

We now assemble all the significant quantities of interest i.e. $q_{2,2}^\xi$, $p_{2,2}^\xi$, $n_{2,2}^\xi$, $|\xi'_{2,2}|$ and $e_{2,2}^{\xi}$ for multiple different initial data configurations in a tabular format: 
\begin{table}[H]
\centering
\begin{tabular}{ || l |c|c|c|c|c || }
\hline
ID & $q_{2,2}^\xi$ & $p_{2,2}^\xi$ & $n_{2,2}^\xi$ & $|\xi'_{2,2}|$ & $e_{2,2}^{\xi}$ \\
\hline
\hline
1.0 & -1.072 & 0.9016 & 1.981 & 17.3263 & 0.0359 \\
\hline
1.1 & -1.072 & 0.9013 & 2.019 & 37.3826  & 0.0785 \\
\hline
1.2 &  -1.071 & 0.9010 & 2.052 & 44.9732 & 0.0958 \\
\hline
1.3 &  -1.071 & 0.9006 & 2.101 & 29.9314 & 0.0652 \\
\hline
1.4 &  -1.070 & 0.8997 & 2.102 & 12.3420 & 0.0273 \\
\hline
\end{tabular}
\caption{The parameters used in the expansion (\ref{ori_expansion2}), and the values of $|\xi'_{2,2}|$ and $e_{2,2}^{\xi}$ for multiple initial data configurations. The table depicts values of those quantities for which the initial Gaussian’s center is at $\rho/M = \{1.0, 1.1, 1.2, 1.3, 1.4\}$.} 
\label{table}
\end{table}
In Fig.~\ref{fig:OriPhi_r}, we plot $g(t,r)\;|\xi_{2,2}(t,r)|$ as a function of $r/M$ where $g(t, r) = (t/M)^2(r/M)^{-1}(r-M)$, as a proxy estimate of $e_{2,2}^{\xi}$. It is clear that this radial dependence is fairly robust for various times $t/M\in[1100, 1600]$, at least for small $r$. 
Finally, to show that we can predictably compute the horizon quadrupolar charge $|\xi'_{2,2}|$ using values of $e_{2,2}^{\xi}$, we plot both these quantities against each other for multiple initial data configurations in Fig.~\ref{fig:HvsE}. We see that the data is well described by a straight line relationship. Fitting our numerical data to $e^{\xi}_{2,2} = \alpha |\xi'_{2,2}| + \beta$, we find that $\alpha = 0.00211 \pm 0.00002$ and $\beta=0.0 \pm 0.0015$. The procedure for calculating the standard errors in the slope and intercept are the same as performed in~\cite{bks23}.

It is worth pointing out that while we have demonstrated that we can compute the proposed non-axisymmetric charge through values of $\xi$ external to the black hole, we currently do not have an explicit and detailed proposal on how to make measurements of $\psi_0$ and $\psi_4$ in a realistic setting using gravitational wave detectors. We leave that open for future work. One comment we can make that is that a detector that would be highly sensitive in the very low-frequency regime (for eg. LISA or similar future detector) would be better to perform such a measurement. 

Note that while eventually one may be able to use such a quadrupolar or octupolar charge to develop a measurable ``signature'' for black hole extremality, there are other known gravitational wave signatures that perhaps are more relevant and interesting in the near term in the context of binary black hole systems. For example, in Ref.~\cite{nek-sign,nek-sign2} the authors propose a signature of extremality based on the quasi-normal ringing phase, i.e. an inverse time decay envelope as opposed to exponential~\cite{nek-sign} and a time-variation in the frequency~\cite{nek-sign2}. In Ref.~\cite{nek-sign3} the authors show that for near extremal black holes, the amplitude of the waveform begins to decrease in the later stages of inspiral and the frequency approaches a fixed value equal to twice the horizon frequency -- a distinct difference from the typical ``chirp'' signal in sub-extremal systems.

\section{Conclusions}
In summary, this work attempts to extend the notion of  ``horizon hair'' to include non-axisymmetric gravitational perturbations on extremal Kerr black hole backgrounds. The introduced expression for a non-axisymmetric charge -- based on the Beetle-Burko radiation scalar, a coordinate and tetrad transformation invariant quantity -- along with the associated Ori-coefficient of this quantity, suggests the potential existence of observable features at a finite distance from the horizon. Our numerical simulations provide strong and reliable evidence supporting the potential existence of this non-axisymmetric, gravitational hair in extremal Kerr spacetime. Note, however, that our proposed expression is only valid at late-times, so it should interpreted as the first term in the inverse-time expansion of a conjectured conserved charge. This work focused on the dominant quadrupolar multipole mode and the subdominant octupolar mode; in future work we plan to study higher-order modes. Our work expands the current knowledge regarding the dynamics of extremal black holes and contributes to the broader comprehension of the nature of gravitational perturbations in these extreme environments. The findings presented in this paper pave the way for further investigations into the observational consequences and implications of non-axisymmetric ``horizon hair'' in extremal Kerr black holes.\\

\noindent{\em Acknowledgements:} 
The authors acknowledge learning about the useful attributes of the Beetle-Burko scalar $\xi$ from Lior Burko and some early analysis performed by Kevin Gonzalez-Quesada. G.K. acknowledges support from NSF Grants No. PHY-2307236 and DMS-2309609. S.B. acknowledges support of NSF grants PHY-2307236 and DMS-2309609. All computations were performed on the UMass-URI UNITY HPC/AI cluster at the Massachusetts Green High-Performance Computing Center (MGHPCC). We also acknowledge the use of {\em ChatGPT} for assistance with summarizing some background material in an initial draft. 

\appendix
\section{Additional Numerical Results}
In this Appendix, we present some secondary numerical results for the non-axisymmetric gravitational Kerr perturbations that are relevant to our construction. 

\begin{figure}[H]
    \centering
    \includegraphics[width=\columnwidth]{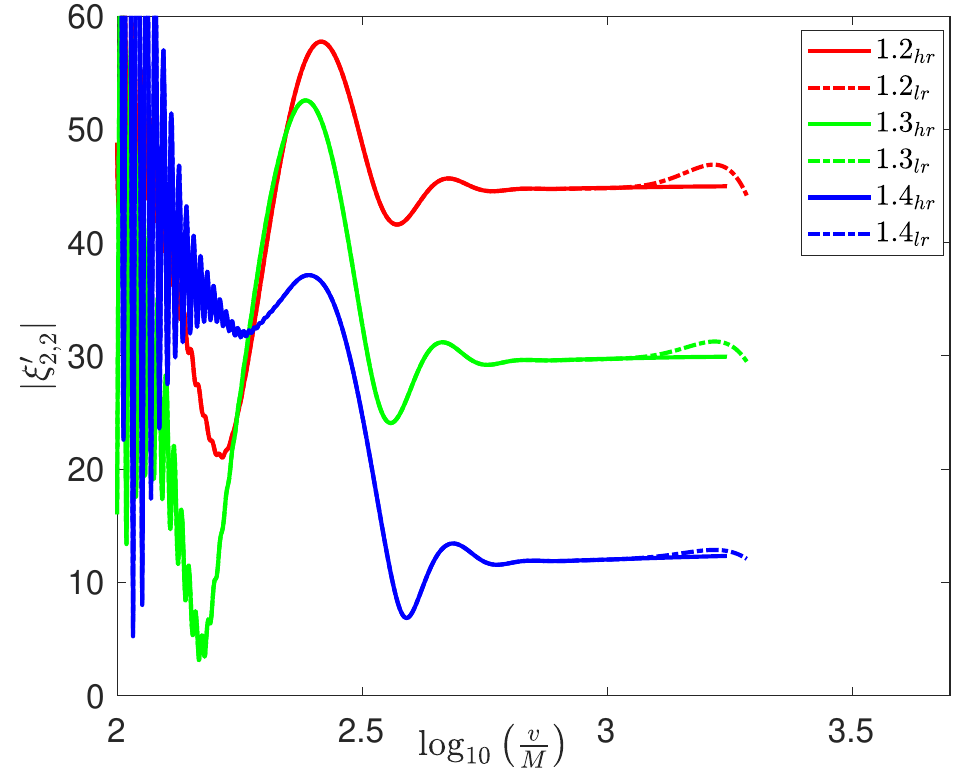}
    \caption{The proposed non-axisymmetric charge $|\xi'_{2,2}|$ on $\ml{H}^+$ as function of advanced time $v$  for two different computational grid densities. The solid lines refer to the higher resolution cases, while the dashed lines refer to the half the grid density of the former. One can see that the numerical results converge reasonably well over the domain of interest.}
    \label{fig:conv}
\end{figure}

\begin{figure}[H]
    \includegraphics[width=\columnwidth,keepaspectratio]{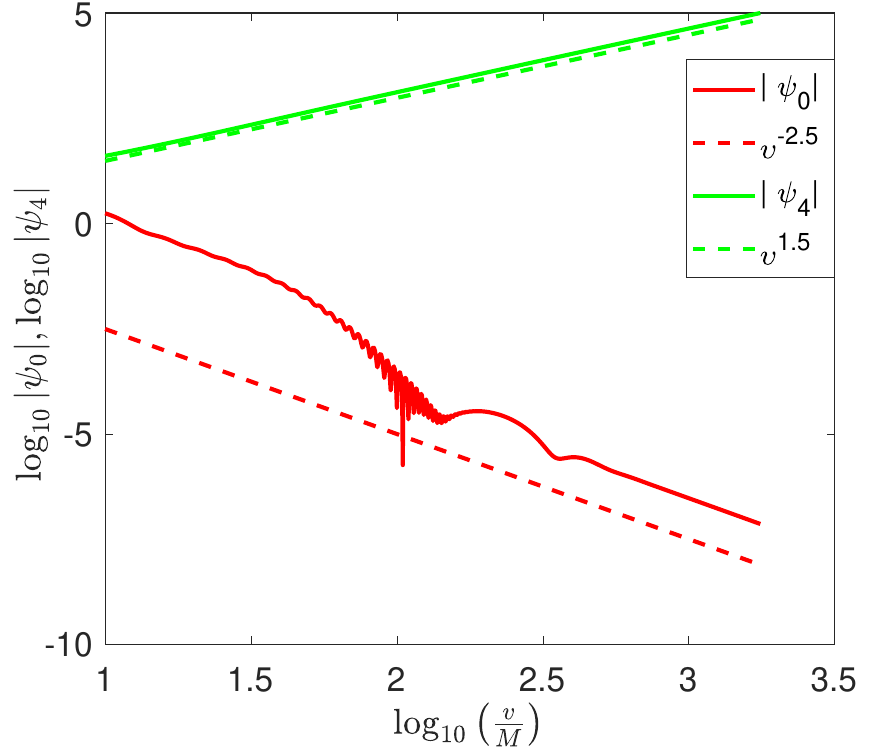}
    \caption{ The behavior of the Weyl scalars, $|\psi_0|$ and $|\psi_4|$ on $\ml{H}^+$ as a function of advanced time $v$.}
    \label{fig:tails}
\end{figure}

\begin{figure}[!h]
    \includegraphics[width=\columnwidth,keepaspectratio]{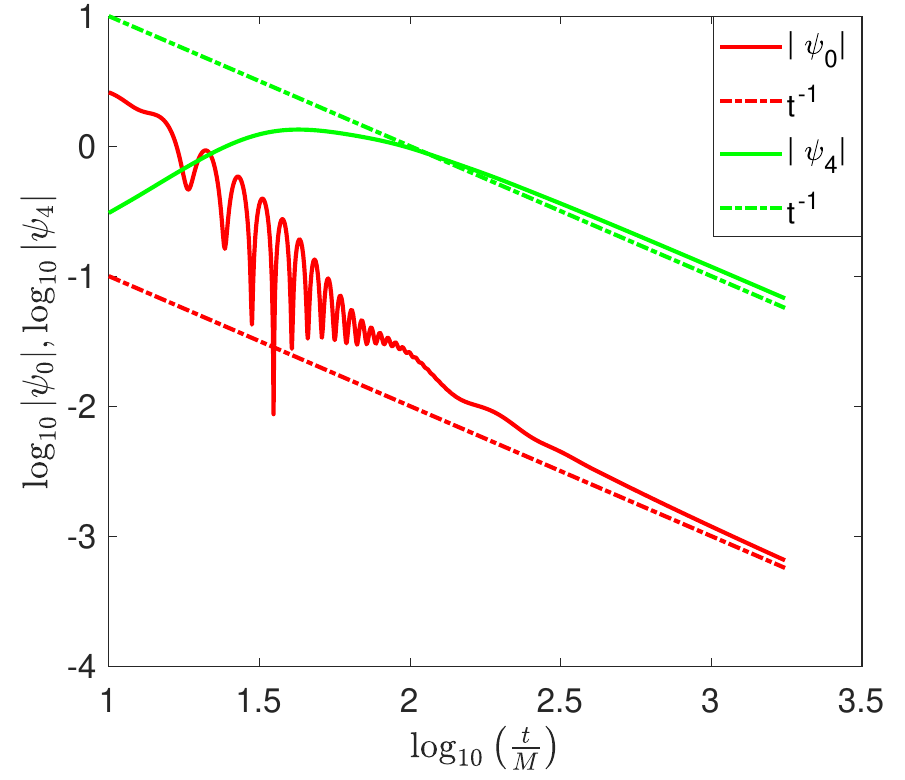}
    \caption{The behavior of the Weyl scalars, $|\psi_0|$ and $|\psi_4|$  at $r=2.235$ as function of time $t$. Note that the fall-off rate in the bulk is the same for both gravitational scalars as argued in~\cite{gz18, casals-2019}.}
    \label{fig:tails2}
\end{figure}

We begin with Fig.~\ref{fig:conv} that presents a numerical convergence test. We depict numerical results for the non-axisymmetric charge $|\xi'_{2,2}|$ for two different numerical resolutions and show that we obtain reasonably good convergence over the duration of interest. The higher resolution data has grid separation $\approx M / 1.5\times10^4$ while the lower resolution case $\approx M / 7.5\times10^3$. 
In Fig.~\ref{fig:tails}, we show the late-time asymptotic power-law decay rates of Weyl scalars $\psi_4$ and $\psi_0$ on $\ml{H}^+$ for $(l, m)=(2,2)$ perturbation. Our numerical results match well with earlier analytical and numerical studies~\cite{bk18, casals-2016}.
Finally, in Fig.~\ref{fig:tails2}, we see the late-time asymptotic power-law decay rates of Weyl scalars $\psi_4$ and $\psi_0$ outside the black hole (i.e., in the bulk) on a time-like curve with a fixed (arbitrarily chosen) value of $r = 2.235$ as a function of Boyer-Lindquist time, $t$. The $1/v$ fall-off has previously been analytically argued in~\cite{gz18, casals-2019}.

\clearpage
\begin{thebibliography}{99}

\bibitem{bek72a} J. D. Bekenstein, Phys. Rev. Lett. 28, 452 (1972)
\bibitem{bek72b} J. D. Bekenstein, Phys. Rev. D 5, 1239 (1972)
\bibitem{bek72c} J. D. Bekenstein, Phys. Rev. D 5, 2403 (1972)

\bibitem{are12} S. Aretakis, arXiv:1206.6598 [gr-qc] (2012)
\bibitem{aag18} Y. Angelopoulos, S. Aretakis, and D. Gajic, Phys. Rev. Lett. 121, 131102 (2018)
\bibitem{lmrt13} J. Lucietti, K. Murata, H.S. Reall, and N. Tanahashi, J. High Energy Phys. 2013, 35 (2013)
\bibitem{zimm17} P. Zimmerman, Phys. Rev. D 95, 124032 (2017)
\bibitem{lr12} J. Lucietti and H.S. Reall, Phys. Rev. D 86, 104030 (2012)
\bibitem{bk18} L.M. Burko and G. Khanna, Phys. Rev. D 97, 061502(R) (2018)
\bibitem{gz18} S.E. Gralla and P. Zimmerman, Class. Quantum Grav. 35, 095002 (2018)
\bibitem{bks21} L. M. Burko, G. Khanna, S. Sabharwal, Phys. Rev. D 103, 021502 (2021)
\bibitem{bks23} L. M. Burko, G. Khanna, S. Sabharwal, Phys. Rev. D 107, 124023 (2023)
\bibitem{aks23} S. Aretakis, G. Khanna, S. Sabharwal, arXiv:2307.03963 (2023)
\bibitem{ori} A. Ori, arXiv:1305.1564 (2013)
\bibitem{gajic} D. Gajic, arXiv:2302.06636 (2023)

\bibitem{nek} L. M. Burko, G. Khanna, S. Sabharwal, Phys. Rev. Research 1, 033106 (2019)
\bibitem{nek2} K. Gonzalez-Quesada, S. Sabharwal, G. Khanna,  Phys. Rev. D 105, 044032 (2022)

\bibitem{bkz16} L. M. Burko, G. Khanna, A. Zenginoğlu, Phys. Rev. D 93, 041501(R) (2016)
\bibitem{z08} A. Zenginoğlu, 2008 Class. Quantum Grav. 25 145002
\bibitem{z11}  A. Zenginoğlu, J. Comput. Phys. 230, 2286 (2011).
\bibitem{zk11} A. Zenginoğlu, G. Khanna, Phys. Rev. X 1, 021017 (2011)
\bibitem{hbnz14} E. Harms, S. Bernuzzi, A. Nagar, and A. Zenginoğlu, Class. Quantum Grav. 31, 245004 (2014)
\bibitem{camc21} S. E. Field, S. Gottlieb, Z. J. Grant, L. F. Isherwood, G. Khanna, Commun. Appl. Math. Comput. (2021)
\bibitem{teuk1973} S. Teukolsky,  Ap. J. {\bf 185}, 635 (1973)
\bibitem{petrov} A.Z.~Petrov, {\it Einstein Spaces} (Pergamon, Oxford, 1969). 
\bibitem{beetle-burko-2002} C.~Beetle and L.M.~Burko, Phys.\ Rev.\ Lett.\ {\bf 89}, 271101 (2002). 
\bibitem{BBBN-2005}  C.~Beetle, M.~Bruni, L.M.~Burko, and A.~Nerozzi,  Phys.\ Rev.\ D {\bf 72}, 024013 (2005).
\bibitem{casals-2016} M.~Casals, S.~E.~Gralla, and P.~Zimmerman, Phys.~Rev.~D {\bf 94}, 064003 (2016). 
\bibitem{casals-2019} M.~Casals and P.~Zimmerman, Phys.~Rev.~D {\bf 100}, 124027 (2019). 
\bibitem{nek-sign} L. M. Burko, G. Khanna, Phys. Rev. D {\bf 94}, 084049 (2016).
\bibitem{nek-sign2} N. E. M. Rifat, G. Khanna, L. M. Burko, Phys. Rev. Research {\bf 1}, 033150 (2019).
\bibitem{nek-sign3} S. E. Gralla, S. A. Hughes, N. Warburton, Class. Quant. Grav. {\bf 33}, 155002 (2016).
\bibitem{Poisson:2004cw}
E.~Poisson,
Phys. Rev. D \textbf{70}, 084044 (2004).
\bibitem{hbh-1}P. Bizon, Phys. Rev. Lett. 64, 2844 (1990).
\bibitem{hbh-2}M.S. Volkov and D.V. Gal’tsov, Sov. J. Nucl. Phys. 51, 1171 (1990).
\bibitem{hbh-3}H. P. Kunzle and A. K. M. Masood-ul-Alam, J. Math. Phys. 31, 928 (1990).
\bibitem{hbh-33}J.A. Smoller, A.G. Wasserman, S.T. Yau and J. McLeod, Comm. Math. Phys. 143 (1991), 115.
\bibitem{hbh-4}P. Bizon and T. Chamj, Phys. Lett. B 297, 55 (1992).
\bibitem{hbh-5}M. Heusler, S. Droz, and N. Straumann, Phys. Lett. B 268, 371 (1991); 271, 61 (1991); 258, 21 (1992).
\bibitem{hbh-6}B. R. Greene, S. D. Mathur, and C. M. O’Neill, Phys. Rev. D 47, 2242 (1993).
\bibitem{hbh-7} B. Kleihaus and J. Kunz, Phys. Rev. Lett. 86, 3704 (2000).
\bibitem{hbh-8} A. Corichi, U. Nucamendi, and D. Sudarsky, Phys. Rev. D 62, 044046 (2000).
\bibitem{hbh-9} N. Dar\'{\i}o, Q. Hernando, and D. Sudarsky, Phys. Rev. Lett. 76, 571 
\bibitem{hbh-10} S. Hod, Phys. Rev. D 84, 124030 (2011).
\bibitem{cbcr} C. Cherubini, D. Bini, S. Capozziello, R. Ruffini, Int. J. Mod. Phys. D, 11(6), 827-841 (2002).
\bibitem{rpm} R. P. Macedo, A. Zenginoglu, Front. Phys. 12, 1497601 (2025).
\bibitem{jlr} J. L. Ripley, Classical Quantum Gravity 39, 1361-6382 (2022).
\bibitem{jms} J. L. Jaramillo, R. P. Macedo, L. A. Sheikh, Phys. Rev. X 11, 031003 (2021).
\bibitem{dmbcj} K. Destounis, R. P. Macedo, E. Berti, V. Cardoso, J. L. Jaramillo, Phys. Rev. D 104, 084091 (2021).
\bibitem{akk} A. Alaee, M. Khuri, H. Kunduri, Phys. Rev. Lett. 119, 071101 (2017).
\bibitem{dain} S. Dain, Phys. Rev. Lett. 96, 101101 (2006).
\bibitem{hhr} S. W. Hawking, G. T. Horowitz, S. F. Ross, Phys. Rev. D 51, 4302–4314 (1995).
\bibitem{sv} A. Strominger, C. Vafa, Phys. Lett. B 379, 99–104 (1996).
\bibitem{cdkktp} P. Claus, M. Derix, R. Kallosh, J. Kumar, P. K. Townsend, A. Van Proeyen, Phys. Rev. Lett. 81, 4553–4556 (1998).
\bibitem{ghw} S. E. Gralla, S. A. Hughes, N. Warburton, Class. Quantum Grav. 33, 155002 (2016).
\bibitem{gls} S. E. Gralla, A. Lupsasca, A. Strominger, Mon. Not. R. Astron. Soc. 475, 3, 3829–3853 (2018).
\bibitem{vmqr} M. Volonteri, P. Madau, E. Quataert, M. Rees, Astrophys. J. 620, 69–77 (2005).
\bibitem{bren} L. Brenneman, et al., Astrophys. J. 736, 103 (2011).

\end{thebibliography}
\end{document}